\documentclass{article}
\usepackage{amsmath}

\input{tcilatex}

\begin{document}

\begin{center}
{\LARGE Reduction and Emergence}

{\LARGE in Bose-Einstein Condensates\bigskip }

{\Large Richard Healey\bigskip }

Philosophy Department, University of Arizona,

213 Social Sciences, Tucson, AZ 85721-0027

rhealey@email.Arizona.edu\bigskip

\textit{Abstract}
\end{center}

A closer look at some proposed \textit{Gedanken}-experiments on BECs
promises to shed light on several aspects of reduction and emergence in
physics. These include the relations between classical descriptions and
different quantum treatments of macroscopic systems, and the emergence of
new properties and even new objects as a result of spontaneous symmetry
breaking.

\section{Introduction}

Not long after the first experimental production of a Bose-Einstein
condensate (BEC) in a dilute gas of rubidium in 1995$^{(1)}$, experiments
demonstrated interference between two such condensates$^{(2)}$. Interference
is a wave phenomenon, and here it was naturally taken to involve a
well-defined phase difference between two coherent matter waves---the BECs
themselves. Experimental phenomena associated with well-defined
phase-differences were already familiar from other condensed matter systems.
The alternating current observed across a Josephson junction between two
similar superconductors was (and is) explained by appeal to their varying
phase-difference induced by a constant voltage difference across the
junction. These two phenomena are now considered manifestations of quantum
behavior at the macroscopic---or at least mesoscopic---level since they
involve very large numbers of atomic or sub-atomic systems acting in
concert, and it is the theory of quantum mechanics that has enabled us to
understand and (at least in the second instance) to predict them, both
qualitatively and in quantitative detail. They are among a variety of
phenomena manifested by condensed matter that have been described as emergent%
$^{(3),(4)}$, in part as a way of contrasting them with phenomena amenable
to a reductive explanation in terms of dynamical laws governing the behavior
of their microscopic constituents.

\qquad While some kind of contrast with reduction is almost always intended
by use of the term `emergent' (or its cognates), the term has been widely
applied to items of many categories on diverse grounds. After briefly
commenting in section 2 on philosophers' attempts to regiment usage, I focus
on a cluster of issues surrounding the emergence of a definite phase in BECs
and related systems.

\qquad It is widely (though not universally) believed that the concept of
broken symmetry is key to understanding not only the Josephson effect and
interference of BECs but also many other phenomena involving condensed
matter.

\qquad When the state of a condensate is represented by a mathematical
object with $U(1)$ symmetry, spontaneous breaking of this symmetry is
associated with a definite phase---the complex argument of an order
parameter such as the expectation-value of a field operator. It may be said
that this phase emerges as a result of such spontaneous symmetry breaking.
Analogies are often drawn between this spontaneously broken phase symmetry
and the breaking of rotational symmetry as the magnetization of a Heisenberg
ferromagnet or the axis of a crystal acquires a definite orientation. But
the attribution of a definite value for the phase of a condensate raises a
thicket of problems that challenge these analogies.

\qquad While the orientation of a crystal or a Heisenberg ferromagnet has
direct operational significance, it is at most the relative phase of two or
more condensates that is manifested in interference experiments: the
absolute phase of a condensate is generally taken to be without physical
significance. A second issue concerns measurements of the relative phase of
condensates. In quantum mechanics, a measurable magnitude (an
\textquotedblleft observable\textquotedblright ) is represented by a
self-adjoint operator, and the possible results of a measurement of this
observable are given by the spectrum of this operator. But there are
powerful reasons for denying that observables generally have values for
measurement to reveal. If the relative phase were represented by such an
operator, then the appearance of a definite (relative) phase on measurement
is no indication of a definite pre-existing phase in the condensate. Rather
than emerging spontaneously, the definite phase would be precipitated by the
measurement itself.

\qquad A number of recent papers have treated the emergence of a definite
relative phase between BECs as a stochastic physical process that occurs as
a result of multiple measurements of quantum observables, each on a
different microscopic constituent of the BECs$^{(5)-(11)}$. The measured
observable is not the phase itself, so there is no need to represent this by
an operator. Indeed, as section 4 explains, the emerging relative phase
plays the role of a kind of \textquotedblleft hidden
variable\textquotedblright\ within a standard quantum mechanical analysis.
This analysis involves no appeal to spontaneous symmetry-breaking. While
some have embellished the analysis by explicit appeal to von Neumann's
controversial projection postulate (\textquotedblleft
collapse\textquotedblright\ of the wave-function on measurement), this
proves unnecessary: all that is required is standard Schr\"{o}dinger quantum
mechanics, including the Born rule for joint probabilities. One way to look
at this quantum mechanical analysis is as a reduction of the theoretical
treatment of relative phase in terms of spontaneous symmetry-breaking. But
this reduction would also involve elimination, in so far as it assumes there
is no well-defined relative phase prior to the measurements that prompt its
emergence.

\qquad A striking feature of the quantum mechanical analysis is that
macroscopic values for observables also emerge in the stochastic process
that produces a well-defined relative phase. These include transverse spin
polarization in a region occupied by two BECs, each composed of particles
with aligned spins, where the two alignments are in opposite directions. The
measurements that induce this macroscopic spin polarization are themselves
microscopic, and may occur in a distant region. As section 5 explains, this
\textquotedblleft nonlocal\textquotedblright\ emergence of macroscopic
values violates expectations based on a common understanding of the
Copenhagen interpretation, and has been presented as a strengthening of
EPR's challenge to that interpretation$^{(12)}$. Section 6 considers a
possible Bohrian response to this challenge and explains why this is in
tension with the common view that the classical features of macroscopic
objects may be derived from quantum theory. This may prompt one to question
the reduction of classical to quantum physics.

\qquad For a global $U(1)$ symmetry, Noether's first theorem implies the
existence of a conserved quantity, which may in this case be identified with
the number of bosons present in a condensate. Broken global $U(1)$ symmetry
then apparently implies a condensate composed of an indeterminate number of
bosons. While coherent laser light has long been accepted as an example of a
condensate with an indeterminate number of massless bosons, an indeterminate
number of atoms in a BEC/Cooper pairs in a superconductor threatens
cherished beliefs about conservation of mass, and baryon/lepton number.
Section 7 addresses the question: Do we have here an emergent object---an
object not composed of any definite number of its constituents?

\qquad The present paper attempts no more than a preliminary survey of a
cluster of complex interrelated issues concerning reduction and emergence in
Bose-Einstein condensates, each of which will repay detailed further study.

\section{Emergence and Reduction}

In physics and elsewhere, reduction and emergence are characteristically
taken to label opposing views of a single relation, but lack of clarity
about the nature of the relation and the identities of the relata often
results in debates between \textquotedblleft
reductionists\textquotedblright\ and their opponents that generate more heat
than light. One problem is that while it is typically phenomena, behavior,
properties, objects, etc. that are said (or denied) to be emergent,
reduction is more commonly thought of as a relation between theories,
theoretical descriptions, sciences or laws (strictly, law statements). So
while emergence is a relation that may or may not hold between items in the
world that scientists study, reduction is a relation applicable only to
products of that study. This division is not hard and fast\footnote{%
In his qualified defense of reductionism, Weinberg$^{(13)}$ casts reduction
and even reductive explanation in ontological rather then epistemological or
methodological terms. He freely admits that a scientist's best strategy in
understanding a phenomenon is often not to look to the fundamental laws that
govern the elementary constituents of the systems involved, even while
maintaining that it is those laws that \textquotedblleft ultimately
explain\textquotedblright\ it.}. But it is a division I shall respect in my
usage in this paper.

\qquad In their attempts to clarify the notion of emergence, philosophers
have typically begun by concentrating their efforts on the emergence of
properties. No consensus has been reached, and a number of alternative
analyses have been proposed$^{(14)-(17)}$. Rather than take these as rival
attempts to state necessary and sufficient conditions for the correct
application of the term `emergent property', one should view them as
alternative explications of the same rough idea---that an emergent property
is one that is somehow autonomous from more elementary underlying structures
out of which it arises. Each may prove useful in marking some contrast that
is important in a different application. One common application of the
notion of emergence is to the mind: philosophers and cognitive scientists
have debated the emergence of consciousness and other mental properties from
underlying physical processes involving the brain. But here I am interested
in contrasting specific physical properties (or, in one case, objects) with
others as to their autonomy from or dependence on more elementary physical
structures.

\qquad The phase of a condensate is the first such property, and the
underlying structures are the properties and arrangement of its constituent
particles. The phase of a condensate is actually a real-valued magnitude,
though any qualitative (i.e. non-numerical) property may be so
regarded---it's values may be taken to be 1 (for possessed) and 0 (for not
possessed). Other magnitudes of systems of condensates may also be
considered emergent, including spin polarization, magnetization and electric
current. We shall see that not one but several senses of emergence turn out
to be usefully applied to these properties.

\qquad Broken symmetries associated with phase transitions in condensates
have been taken to give rise to emergent phenomena by both physicists and
philosophers$^{(3),(4),(16),(18)}$\footnote{%
Though Anderson$^{(18)}$ doesn't use the word `emergent'. It is an
unfortunate linguistic accident that in the expression `phase transition'
the word `phase' refers to states of matter themselves (e.g.
superconducting), not to the complex argument of a parameter that may be
used to characterize their degree of order.}. Weinberg$^{(19)}$ even defines
a superconductor as \textquotedblleft simply a material in which
electromagnetic gauge invariance is spontaneously broken\textquotedblright .
This at least suggests that it is spontaneously broken symmetry that marks
properties of matter as emergent in a novel phase. If so, properties of
matter in that phase that can be accounted for without appealing to broken
symmetry would not count as emergent.

\qquad In one sense, emergence is a diachronic process rather than a
synchronic condition. Phase transitions occur as dynamical processes,
whether or not the symmetry of the prior state is physically broken during
this process. So a phase of matter with striking properties may emerge
dynamically even though these properties are not sufficiently autonomous
from the underlying structure in the new phase to count as (synchronically)
emergent.

\qquad I think there is another possible use of `emergent', as applied to
properties of a complex system which is, perhaps, illustrated by the
emergence of a definite (relative) phase in BECs. Consider such
\textquotedblleft sensory\textquotedblright\ predicates as red, malodorous,
bitter, silky or even wet or hard\footnote{%
See Wilson's$^{(20)}$ extended exploration of the sensory concomitants of
the first and last of these terms and their bearing on the character of any
corresponding property.}. In paradigm cases, though certainly not always,
these are applied to a macroscopic object on the basis of the response it
elicits in a human who interacts with that object in a minimally invasive
way---unfortunately, looking at a red traffic light is not an effective way
to turn it green, and nor does sniffing rotten meat improve its smell. But
do such predicates pick out a corresponding property of that object?

\qquad Many and varied answers to that question have been proposed
throughout the history of philosophy and natural science. Some have defended
a positive answer by claiming that a property such as the redness of an
object supervenes on more fundamental properties of the microscopic
constituents of that object that are not themselves red. Others have denied
the existence of any property of redness, flushed with the prospect of a
complete scientific explanation of our ability to perceive, classify and
reliably communicate about those things we call red based only on their
fundamental microphysical properties and ours. Philosophical accounts of
emergence generally presuppose that emergent properties are real, even if
they supervene on an underlying microphysical basis. But if one had a
complete scientific explanation of our ability to perceive, classify and
reliably communicate about those things we call red, that might itself be
offered as an account of the emergence of redness even if there is no such
property! For the account would explain the success of our common practice
of calling things red and so license the continuance of that practice.

\section{BEC Phase as Emerging from Spontaneous Symmetry Breaking?}

In his seminal essay Anderson$^{(18)}$ takes the general theory of broken
symmetry to offer an illuminating formulation of how the shift from
quantitative to qualitative differentiation characteristic of emergence
takes place\footnote{%
\textquotedblleft at each level of complexity entirely new properties
appear\textquotedblright\ ($(18)$, p.393).}. In agreement with Weinberg$%
^{(19)}$ he mentions superconductivity as a spectacular example of broken
symmetry, though he gives several others.

\begin{quote}
\qquad The essential idea is that in the so-called $N\rightarrow \infty $
limit of large systems (on our own, macroscopic scale) it is not only
convenient but essential to realize that matter will undergo sharp, singular
\textquotedblleft phase transitions\textquotedblright\ to states in which
the microscopic symmetries, and even the microscopic equations of motion,
are in a sense violated. (\textit{op}. \textit{cit}. p.395)
\end{quote}

After the 1995 experimental production of BECs in dilute gases, Laughlin and
Pines$^{(3)}$ were able to add \textquotedblleft the newly discovered atomic
condensates\textquotedblright\ as examples that display emergent physical
phenomena regulated by higher organizing principles. Since they cite
Anderson's paper approvingly and take a principle of continuous symmetry
breaking to explain (the exact character of) the Josephson effect, it is
reasonable to conjecture that they would join Anderson in taking the phase
transition from a normal dilute gas to a BEC as well as that from a normal
metal to a superconducting state to involve spontaneous symmetry breaking.

\qquad What symmetry is taken to be broken in the transition to the
condensed phase of a BEC? The transition is from a less to a more ordered
state, whose order may be represented by a so-called order parameter.
According to Leggett$^{(21)}$ (p. 38) the order parameter characterizing a
BEC (especially in the case of dilute gases including rubidium) is often
taken to be a complex-valued function---the expectation value of a Bose
field operator in the given quantum state.%
\begin{equation}
\Psi \left( \mathbf{r,}t\right) =\left\langle \hat{\psi}\left( \mathbf{r,}%
t\right) \right\rangle
\end{equation}%
If this is written as%
\begin{equation}
\Psi \left( \mathbf{r,}t\right) =\left\vert \Psi \left( \mathbf{r,}t\right)
\right\vert e^{i\varphi \left( \mathbf{r,}t\right) }
\end{equation}%
then the phase $\varphi \left( \mathbf{r,}t\right) $ parametrizes an element
of the group $U(1)$. If the equations describing the field of the condensate
are symmetric under global $U(1)$ transformations, then changing the order
parameter by addition of an arbitrary constant to the phase will take one
solution into a distinct solution. Global $U(1)$ symmetry will be broken by
choice of one such value.

\qquad An analogy is often drawn to the broken rotation symmetry of the
Heisenberg ferromagnet as the spins of all its magnetic dipoles align along
some arbitrary direction in the ground state. That fits Anderson's quoted
description well, since the phase transition to one such highly ordered
ground state of the ferromagnet is a good example of the kind of
spontaneously broken symmetry amenable to idealized treatment as a quantum
system with an infinite number of degrees of freedom\footnote{%
See, for example, Ruetsche$^{(22)}$.}. In contrast to the case of a quantum
system with a finite number of degrees of freedom, degenerate ground states
of such a system cannot generally be superposed to give another state since
they appear in distinct, unitarily inequivalent, representations of the
fundamental commutation relations. Spontaneous breaking of the rotational
symmetry of the Heisenberg ferromagnet corresponds to the adoption of one
out of the many states in which the dipoles of the ferromagnet are all
aligned. In two or more dimensions, this means breaking of a continuous
rotational symmetry. By Goldstone's theorem$^{(23)}$, when such a continuous
symmetry is broken in quantum mechanics the Hamiltonian has no energy gap%
\footnote{%
As Streater$^{(24)}$ proved for the Heisenberg ferromagnet: this gives rise
to the possibility of spin waves of arbitarily small energy.}: in a quantum
field theory this implies the existence of massless Goldstone bosons.

\qquad Pursuing this analogy, spontaneous breaking of the continuous $U(1)$
phase symmetry of a BEC's order parameter could be represented by an
idealized model in which the number of constituent particles is taken to be
infinite, but the density of the BEC is fixed at some low value $\rho $ by
taking the so-called thermodynamic limit $N\rightarrow \infty ,$ $%
V\rightarrow \infty ,$ $N/V=\rho $ (a constant). Then adoption of a definite
phase by a BEC would be an instance of the same kind of spontaneous symmetry
breaking as adoption of a definite direction of magnetization by a
Heisenberg ferromagnet. But there are problems with this analogy, as Leggett$%
^{(21),(25),(26)}$ has noted.

\qquad When rotation symmetry of a Heisenberg ferromagnet is spontaneously
broken, the spins of its components are all aligned along a particular
direction in space. This direction may be operationally defined in many ways
having nothing to do with spin or magnetization: in particular, it need not
be defined in relation to other Heisenberg ferromagnets, either actual or
hypothetical. On the other hand, if the \textit{U(1)} global phase symmetry
of a BEC were to be spontaneously broken, its overall phase would become
well defined only relative to some other BEC of the same kind (for example,
a similarly condensed dilute gas of rubidium 87). At most, a definite phase
consequent upon spontaneously broken symmetry would seem to be an emergent
relational property (cf. Teller$^{(27)}$) of a BEC. Moreover, difficulties
in implementing multiple pairwise phase comparisons between similar BECs
that have never been in contact threaten at least the operational
significance even of such a relational property. Leggett$^{(26)}$ argues
that, at least in the case of superconducting BECs, operational pairwise
phase comparisons among several such BECs will fail to be transitive (though
compare Leggett$^{(28)}$).

\qquad A second problem arises from the need to take the thermodynamic limit
to treat the emergence of relative phase in BECs as an instance of
spontaneous symmetry breaking. No massive BEC system is composed of an
infinite number of elementary bosons. Moreover, while the number of
elementary dipoles in a macroscopic magnet will typically at least be
extremely large (of the order of $10^{23}$ ), the first dilute gas BECs
contained only a few thousand atoms, and even now experimental realizations
have increased this number by only a few factors of $10$. If it were
essential to assume that an \textit{infinite} number of atoms is present in
each of two interfering BECs to explain their observed interference (as the
quote from Anderson might lead one to believe), then one may legitimately
query the value of the explanation. But in fact one need not treat the
emergence of relative phase here as a case of spontaneous symmetry breaking
in the thermodynamic limit, as analyses by Castin and Dalibard$^{(6)}$ and
several subsequent authors have shown.

\qquad In the context of an idealized model of two trapped condensates of
the same atomic species, Castin and Dalibard$^{(6)}$ showed two things:

(1)\qquad No measurements performed on the condensates can allow one to
distinguish between two different quantum representations of this system: By
a uniform average over the unknown relative phase of two coherent states;
and by a Poissonian statistical mixture of Fock states.

(2)\qquad Two different points of view on a system are available: Assuming
an initial pair of coherent states with a definite relative phase,
successive measurements \textquotedblleft reveal\textquotedblright\ that
pre-existing phase in an interference phenomenon; assuming each condensate
is initially\ represented by a definite Fock state, with no well-defined
relative phase, the same sequence of measurements progressively
\textquotedblleft builds up\textquotedblright\ a relative phase between the
condensates as the interference phenomenon is generated.

They take the results of their analysis to show that the notion of
spontaneously broken phase symmetry is not indispensable in understanding
interference between two condensates. I won't explain how they arrived at
these conclusions, since the next section outlines a closely related
analysis by Lalo\"{e} of a similar \textit{Gedankenexperiment} that will
provide a focus for the subsequent discussion. I will merely comment that
Castin and Dalibard$^{(6)}$ assume that the measurements referred to in (2)
are performed in a well-defined temporal sequence on individual elements of
the system of condensates, and that each leaves the rest of the system in
the quantum state it would be assigned if the effect of that measurement
were represented by von Neumann's projection postulate.

\section{The Appearance of Phase Without Symmetry-Breaking}

In 2005 Lalo\"{e}$^{(7)}$ began to develop an elegant framework for
analyzing the emergence of phase in systems of BECs. One important
application is to a system of two BECs, each composed of non-interacting
bosons, and each initially represented by a Fock state corresponding to a
definite number of particles. This provides a simplified and idealized model
for the kind of experimental situation realized by Andrews \textit{et}. 
\textit{al}.$^{(2)}$ that first demonstrated interference between two BECs.
An extension of that model is to measurements on BECs in different internal
states---most simply, each in one of two different one-particle \textit{z}%
-spin states. This enables one to consider the BECs to be initially separate
systems no matter what their spatial overlap: and it naturally suggests the
possibility of a variety of different kinds of measurement capable of
revealing interference between them---of spin-component in any direction in
the \textit{x-y} plane. Such measurements are considered in Mullin, Krotkov
and Lalo\"{e}$^{(8)}$, Lalo\"{e}$^{(12)}$, and Lalo\"{e} and Mullin$^{(10)}$%
: here I follow Lalo\"{e}'s$^{(12)}$ presentation.

\qquad Consider a pair of noninteracting spin-polarized BECs in the
normalized Fock state%
\begin{equation}
\left\vert \Phi \right\rangle =\frac{1}{\sqrt{N_{a}!N_{b}!}}\hat{a}\dag
_{u_{a},\alpha }^{N_{a}}\hat{a}\dag _{v_{b},\beta }^{N_{b}}\left\vert
0\right\rangle  \label{double Fock}
\end{equation}%
representing $N_{a}$ particles with internal ($z$-spin) state $\alpha $ and
spatial state $u_{a}$ and $N_{b}$ particles with orthogonal internal ($z$%
-spin) state $\beta $ and spatial state $v_{b}$, where $\left\vert
0\right\rangle $ is the vacuum state.

If $\hat{\Psi}_{\alpha }(\mathbf{r})$ is the field operator for $z$-spin $%
\alpha $, $\hat{\Psi}_{\beta }(\mathbf{r})$ for $z$-spin $\beta $, and $%
^{\dag }$ indicates the adjoint operation, \ then the number density
operator of the BECs is%
\begin{equation}
\hat{n}(\mathbf{r})=\hat{\Psi}_{\alpha }^{\dag }(\mathbf{r})\hat{\Psi}%
_{\alpha }(\mathbf{r})+\hat{\Psi}_{\beta }^{\dag }(\mathbf{r})\hat{\Psi}%
_{\beta }(\mathbf{r})  \label{number density}
\end{equation}%
and the density operator for their spin component in a direction in the $x-y$
plane at an angle $\varphi $ from the $x$-axis is%
\begin{equation}
\hat{\sigma}_{\varphi }(\mathbf{r})=e^{-i\varphi }\hat{\Psi}_{\alpha }^{\dag
}(\mathbf{r})\hat{\Psi}_{\beta }(\mathbf{r})+e^{+i\varphi }\hat{\Psi}_{\beta
}^{\dag }(\mathbf{r})\hat{\Psi}_{\alpha }(\mathbf{r})  \label{spin density}
\end{equation}%
Suppose that one measurement is made of the $\varphi $-component of particle
spin in a small region of space $\Delta r$ centered around point $\mathbf{r}$%
. The corresponding spin operator is%
\begin{equation}
\hat{S}(\mathbf{r},\varphi )=\int_{\Delta r}d^{3}\mathbf{r}^{\prime }\hat{%
\sigma}_{\varphi }(\mathbf{r}^{\prime })
\end{equation}%
For sufficiently small $\Delta r$, this has only three eigenvalues $\eta
=0,\pm 1$ since no more than one particle would be found in $\Delta r$. The
single-particle eigenstates for finding a particle there with $\eta =\pm 1$
are%
\begin{equation}
\left\vert \Delta r,\eta \right\rangle =\left\vert \Delta r\right\rangle
\otimes \frac{1}{\sqrt{2}}\left[ e^{-i\varphi /2}\left\vert \alpha
\right\rangle +e^{+i\varphi /2}\left\vert \beta \right\rangle \right]
\end{equation}%
where $\left\vert \Delta r\right\rangle $ is a single-particle spatial state
whose wave-function equals $1$ inside $\Delta r$ but $0$ everywhere outside $%
\Delta r$. The corresponding $N$-particle projector is%
\begin{equation}
\hat{P}_{\eta =\pm 1}(\mathbf{r,}\varphi )=\frac{1}{2}\int_{\Delta r}d^{3}%
\mathbf{r}^{\prime }\left[ \hat{n}(\mathbf{r}^{\prime })+\eta \hat{\sigma}%
_{\varphi }(\mathbf{r}^{\prime })\right]  \label{projectors}
\end{equation}%
and the projector for finding no particle there is%
\begin{equation}
\hat{P}_{\eta =0}(\mathbf{r})=\left( \mathbf{1}-\int_{\Delta r}d^{3}\mathbf{r%
}^{\prime }\hat{n}(\mathbf{r}^{\prime })\right)
\end{equation}%
As $\Delta r\rightarrow 0$, the corresponding eigenstates (for variable $%
\mathbf{r}$) form a quasi-complete basis for the $N$-particle space.

Now consider a sequence of $m$ measurements of transverse spin-components $%
\varphi _{j}$ in very small non-overlapping regions $\Delta r_{j}$, each of
volume $\Delta $, centered around points $\mathbf{r}_{j\text{ }}(1\leq j\leq
m)$. Since the projectors for non-overlapping regions commute, the joint
probability for detecting $m$ particles with spins $\eta _{j}$ in regions $%
\Delta r_{j}$ is%
\begin{equation}
\left\langle \Phi \left\vert \hat{P}_{\eta _{1}}(\mathbf{r}_{1}\mathbf{,}%
\varphi _{1})\times \hat{P}_{\eta _{2}}(\mathbf{r}_{2}\mathbf{,}\varphi
_{2})\times ...\times \hat{P}_{\eta _{m}}(\mathbf{r}_{m}\mathbf{,}\varphi
_{m})\times \right\vert \Phi \right\rangle  \label{joint probabilities}
\end{equation}%
Using (\ref{projectors}) together with (\ref{number density}) and (\ref{spin
density}) this gives a product of several terms, each containing various
products of field operators. Since these commute, we can push all the
creation operators to the left and all the annihilation operators to the
right. Expanding the field operators in terms of a basis $\left\vert
u_{a},\alpha \right\rangle $, $\left\vert v_{b},\beta \right\rangle $ of
single particle states%
\begin{equation}
\hat{\Psi}_{\alpha }(\mathbf{r})=u_{a}(\mathbf{r})\times \hat{a}%
_{u_{a},\alpha }+...\text{ \ \ ; \ \ }\hat{\Psi}_{\beta }(\mathbf{r})=v_{b}(%
\mathbf{r})\times \hat{a}_{v_{b},\beta }+...
\end{equation}%
But none of the "dotted" terms will contribute to (\ref{joint probabilities}%
), since $\left\vert \Phi \right\rangle $ contains no particles in states
other than $\left\vert u_{a},\alpha \right\rangle $, $\left\vert v_{b},\beta
\right\rangle $.

Each term now contains between $\left\langle \Phi \right\vert $ and $%
\left\vert \Phi \right\rangle $ a string of creation operators followed by a
string of annihilation operators. If a state $\left\vert u_{a},\alpha
\right\rangle $ or $\left\vert v_{b},\beta \right\rangle $ does not appear
exactly the same number of times in each of these, it will not contribute to
(\ref{joint probabilities}): if it does appear exactly the same number of
times in each of these, every creation or annihilation operator will
introduce a factor $\sqrt{N_{a,b}-q}$ where $q$ depends on the term but $q<m$%
. If $m\ll N_{a},N_{b}$, these factors can be approximated by $\sqrt{N_{a,b}}
$ respectively. So now each field operator has been replaced in (\ref{joint
probabilities}) by a factor $\sqrt{N_{a,b}}$ multiplying a position
wave-function $u_{a}$ or $v_{b}$ (or its complex conjugate). But we still
have to take account of particle number conservation in each sequence. This
can be done by a clever trick, using the mathematical identity%
\begin{equation}
\dint\limits_{0}^{2\pi }\frac{d\Lambda }{2\pi }e^{in\Lambda }=\delta _{n,0}
\end{equation}%
By multiplying each $\hat{\Psi}_{\alpha }(\mathbf{r})$ (or rather $\sqrt{%
N_{a}}u_{a}(\mathbf{r})$) by $e^{i\Lambda }$, and each $\hat{\Psi}_{\alpha
}^{\dag }(\mathbf{r})$ (or rather $\sqrt{N_{a}}u_{a}^{\ast }(\mathbf{r})$)
by $e^{-i\Lambda }$, and integrating $\Lambda $ over $2\pi $ (and similarly
for the $b$ particles), we automatically take account of particle number
conservation!

Since $\Delta $ is very small, the spatial wave-functions are each
approximately constant over each region $\Delta r_{j}$.The joint probability
for detection of $m$ particles with spins $\eta _{j}$ in regions $\Delta
r_{j}$ $(1<j<m)$, each of volume $\Delta $, is then 
\begin{equation}
\frac{\Delta ^{m}}{N_{a}!N_{b}!}\dint\limits_{0}^{2\pi }\frac{d\Lambda }{%
2\pi }\dprod\limits_{j=1}^{m}\left\{ 
\begin{array}{c}
N_{a}\left\vert u_{a}(\mathbf{r}_{j})\right\vert ^{2}+N_{b}\left\vert v_{b}(%
\mathbf{r}_{j})\right\vert ^{2}+ \\ 
\eta _{j}\sqrt{N_{a}N_{b}}\left( e^{i\left( \Lambda -\varphi _{j}\right)
}u_{a}(\mathbf{r}_{j})v_{b}^{\ast }(\mathbf{r}_{j})+c.c\right)%
\end{array}%
\right\}
\end{equation}%
or, with $\xi (\mathbf{r})=\arg \left[ u_{a}(\mathbf{r})/v_{b}(\mathbf{r})%
\right] $,%
\begin{equation}
\frac{\Delta ^{m}}{N_{a}!N_{b}!}\dint\limits_{0}^{2\pi }\frac{d\Lambda }{%
2\pi }\dprod\limits_{j=1}^{m}\left\{ 
\begin{array}{c}
N_{a}\left\vert u_{a}(\mathbf{r}_{j})\right\vert ^{2}+N_{b}\left\vert v_{b}(%
\mathbf{r}_{j})\right\vert ^{2}+ \\ 
2\eta _{j}\sqrt{N_{a}N_{b}}\left\vert u_{a}(\mathbf{r}_{j})\right\vert
\left\vert v_{b}(\mathbf{r}_{j})\right\vert \cos (\Lambda +\xi (\mathbf{r}%
_{j})-\varphi _{j})%
\end{array}%
\right\}  \label{prob}
\end{equation}%
Since we are interested in comparing relative probabilities of detection,
the prefixed normalization terms can now be dropped.

First consider the case $m=1$: measurement of $\varphi $-spin on a single
particle. If $\Lambda $ were fixed, these relative detection probabilities
would be just what one would expect from a state with a definite relative
phase between the two condensates, namely the (unnormalized) phase state%
\begin{equation}
\left\vert \Lambda \right\rangle =\left[ N_{a}^{%
{\frac12}%
}\hat{a}\dag _{u_{a},\alpha }e^{-i\Lambda /2}+N_{b}^{%
{\frac12}%
}\hat{a}\dag _{v_{b},\beta }e^{i\Lambda /2}\right] ^{N_{a}+N_{b}}\left\vert
0\right\rangle  \label{phase state}
\end{equation}%
with one-particle pure spin-state density matrix $W$ proportional to%
\begin{equation}
\begin{pmatrix}
N_{a}\left\vert u_{a}(\mathbf{r}_{1})\right\vert ^{2} & \sqrt{N_{a}N_{b}}%
e^{-i\Lambda }u_{a}^{\ast }(\mathbf{r}_{1})v_{b}(\mathbf{r}_{1}) \\ 
\sqrt{N_{a}N_{b}}e^{i\Lambda }u_{a}(\mathbf{r}_{1})v_{b}^{\ast }(\mathbf{r}%
_{1}) & N_{b}\left\vert v_{b}(\mathbf{r}_{1})\right\vert ^{2}%
\end{pmatrix}%
\end{equation}%
Moreover, the expectation value of $z$-spin in state $W$ is proportional to $%
N_{a}\left\vert u_{a}(\mathbf{r}_{1})\right\vert ^{2}-N_{b}\left\vert v_{b}(%
\mathbf{r}_{1})\right\vert ^{2}$, also just what one would expect if $%
\left\vert \Lambda \right\rangle $ represented the definite, preexisting
relative phase between the condensates. But the uniform integral over $%
\Lambda $ "washes out" the appearance of any definite phase relation between
the two condensates, so the overall probability distribution for measurement
of $\varphi $-spin on a single particle corresponds to no interference.

Now consider the case $m=2$: joint measurement of $\varphi _{1}$-spin and $%
\varphi _{2}$-spin with results $\eta _{1},\eta _{2}$ respectively on two
particles. The $\Lambda $-probability distribution for result $\eta _{2}$ 
\textit{conditional on outcome }$\eta _{1}$ is now weighted by a factor that
depends both on the angle $\varphi _{1}$ of the measurement on particle $1$
and on its outcome and location $\left( \eta _{1},\mathbf{r}_{1}\right) $
and is proportional to%
\begin{equation}
N_{a}\left\vert u_{a}(\mathbf{r}_{1})\right\vert ^{2}+N_{b}\left\vert v_{b}(%
\mathbf{r}_{1})\right\vert ^{2}+2\eta _{1}\sqrt{N_{a}N_{b}}\left\vert u_{a}(%
\mathbf{r}_{1})\right\vert \left\vert v_{b}(\mathbf{r}_{1})\right\vert \cos
(\Lambda +\xi (\mathbf{r}_{1})-\varphi _{1})
\end{equation}%
This may well already give rise to a slight correlation between the results $%
\eta _{1},\eta _{2}$: if $\eta _{1}$ is $+1$ and $\varphi _{1}$ and $\varphi
_{2}$ are close, then $\eta _{2}$ is more likely than not also to equal $+1$%
. But as one considers additional transverse spin measurements, strong
correlations become apparent. The probability distribution for the
transverse spin of the $\left( m+1\right) $st particle \textit{conditional
on outcomes }$\eta _{j}$ \textit{for the other} \textit{m} \textit{%
measurements } becomes strongly peaked as $m$ increases. Lalo\"{e}$^{(12)}$
comments

\begin{quote}
When more and more spin measurements are obtained, the $\Lambda $%
-distribution becomes narrower and narrower....Standard quantum mechanics
considers that $\Lambda $ has no physical existence at the beginning of the
series of measurements, and that its determination is just the result of a
series of random perturbations of the system introduced by the measurements.
Nevertheless $\left( \ref{prob}\right) $ shows that all observations are
totally compatible with the idea of a pre-existing value of $\Lambda $ which
is perfectly well defined but unknown, remains constant, and is only
revealed (instead of created) by the measurements. (p. 43)
\end{quote}

It is tempting to think of the emergence of a definite phase here as a
stochastic, dynamical process in which each subsequent transverse
spin-measurement (with increasing probability) renders the relative phase of
the two condensates more definite. But the parameter $\Lambda $ enters the
above quantum mechanical analysis only as a convenient mathematical device
for calculating conditional probabilities, and not (as in $\left( \ref{phase
state}\right) $) as a way of characterizing the state of the condensates
themselves. Moreover, the analysis nowhere appealed to the evolution of the
state of the condensates, whether unitary (in accordance with the Schr\"{o}%
dinger equation) or non-unitary (in accordance with von Neumann's projection
postulate). Though if one were instead to assume a temporal sequence of
projective $\varphi _{j}$-spin measurements, then (neglecting Schr\"{o}%
dinger evolution) the state of the remaining condensate would progressively
come to approximate a state of definite phase.

\section{A Strengthened EPR argument?}

Lalo\"{e}$^{(12)}$ goes on to consider alternative spatial wave-functions $%
u_{a},v_{b}$ for the two condensates. He takes one such configuration to
justify this claim in the abstract to his paper.

\begin{quote}
We study in this article how the EPR argument can be transposed to this
case, and show that the argument becomes stronger, mostly because the
measured systems themselves are now macroscopic. (p.35)
\end{quote}

He makes the simplifying assumption that $u_{a},v_{b}$ have the same phase
at each point $\mathbf{r}$ (though their amplitudes may differ) so $\xi =0$.
He then considers an arrangement in which $u_{a},v_{b}$ overlap only in two
distant regions $A,B$ and $m$ successful transverse spin-measurements are
considered in non-overlapping small regions of $A$. The foregoing analysis
shows that, for $m\sim 100$, the conditional probabilities for the outcomes
of additional transverse spin-measurements \textit{in small non-overlapping
small regions of }$B$ will differ little from corresponding unconditional
probabilities for a phase state with some definite $\Lambda $ (whose value
depends on the outcomes of the $m$ measurements \textit{in }$A$). In
particular, if $B$ contains a macroscopic number of particles there will be
some angle $\varphi _{\Lambda }$ such that each of, say, $10^{23}$
successful individual measurements of $\varphi _{\Lambda }$-spin on
particles in $B$ is almost certain to give outcome $\eta _{\Lambda }=+1$,
conditional on the outcomes of the $m$ measurements in $A$. Lalo\"{e}$%
^{(12)} $ comments

\begin{quote}
Here we have a curious case where it is the measured system itself that
spontaneously creates a pointer made of a macroscopic number of parallel
spins. Moreover, for condensates that are extended in space, ...this process
can create instantaneously parallel pointers in remote regions of space, a
situation obviously reminiscent of the EPR argument in its spin version
given by Bohm. (p.37)
\end{quote}

Recall Bohm's spin version of the EPR \textit{Gedankenexperiment}, featuring
two spin \textit{%
$\frac12$
}particles in the spin singlet state%
\begin{equation}
\left\vert \psi _{s}\right\rangle =\frac{1}{\sqrt{2}}\left[ \left\vert
\uparrow \downarrow \right\rangle -\left\vert \downarrow \uparrow
\right\rangle \right]
\end{equation}%
where $\uparrow (\downarrow )$ labels a $z$-spin eigenvector with positive
(negative) eigenvalue and the order of the arrows in each component of the
superposition corresponds to that of the particles' state spaces. Transposed
to this situation, the intended conclusion of the EPR argument is that
quantum mechanical description is incomplete since the state $\left\vert
\psi _{s}\right\rangle $ does not describe certain "elements of reality"
associated with each of these two particles: for each direction, one such
"element of reality" corresponds to the (eigen)value of spin-component in
that direction which a well-conducted measurement of that spin component
would reveal, were one to be carried out.

After removing the excess erudition of which Einstein complained right after
its publication, the original EPR argument went like this\footnote{%
In a letter to Schr\"{o}dinger of June 19th, 1935 Einstein said that the
main point was, so to speak, overwhelmed by erudition ("die Hauptsache ist,
sozusagen, durch Gelehrsamkeit versch\"{u}ttet").}. Einstein, Podolsky and
Rosen$^{(29)}$ assumed the following sufficient criterion for reality

\begin{quote}
If, without in any way disturbing a system, we can predict with certainty
(i.e. with probability equal to unity) the value of a physical quantity,
then there exists an element of physical reality corresponding to this
quantity. (p.777)
\end{quote}

Suppose one were to get outcome $\eta $ in a measurement of the (arbitrary) $%
\varphi $-component of spin on particle $a$ of a Bohm-EPR\ pair in spin
state $\left\vert \psi _{s}\right\rangle $ at a time when their spatial
state corresponds to $a,b$ being far apart (with negligible probability of
finding $a$ outside $A$ or $b$ outside $B$, where $A,B$ are widely separated
spatial regions). Assuming this is a projective measurement, the resulting
state is%
\begin{equation}
\left\vert \psi ^{\prime }\right\rangle =\left\vert \varphi _{\eta }\varphi
_{-\eta }\right\rangle
\end{equation}%
Applying the Born rule to $\left\vert \psi ^{\prime }\right\rangle $\ one
could predict with probability unity that a (well-conducted) measurement of
the value of the $\varphi $-component of spin on particle $b$ would yield
outcome $-\eta $. EPR\ further take state $\left\vert \psi ^{\prime
}\right\rangle $ to describe $b$ as certainly (with probability unity) 
\textit{having} value $-\eta $ for its $\varphi $-component of spin even if
no measurement is performed on $b$. Assuming (locality) that such a
hypothetical measurement on $a$ alone would not disturb $b$, they infer that
in the hypothetical situation in which (only) $a$ is measured with result $%
\eta $, the $\varphi $-component of spin of $b$ would be $-\eta $, prior to
and independent of the measurement on $a$:\ similarly, in the hypothetical
situation in which (only) $a$ is measured with result $-\eta $, the $\varphi 
$-component of spin of $b$ would be $\eta $, prior to and independent of the
measurement on $a$. Hence in any hypothetical situation in which (only) $a$
is measured, $b$ would have had a definite (though as yet unknown) $\varphi $%
-component of spin, prior to and independent of the measurement on $a$. It
follows that $b$ always \textit{has }a definite (though unknown) $\varphi $%
-component of spin in the spin singlet state, irrespective of what
measurement (if any) one contemplates performing on $a$ or $b$: by symmetric
reasoning, so too does $a$. Together with EPR's necessary condition for
completeness ("every element of the physical reality must have a counterpart
in the physical theory") this establishes the incompleteness of quantum
mechanical description.

Lalo\"{e}$^{(12)}$ takes his BEC \textit{Gedankenexperiment} to strengthen
the EPR argument "mostly because the measured systems themselves are now
macroscopic". In evaluating this claim later, it will be helpful to bear in
mind a more straightforward extension of the Bohm-EPR scenario to the
macroscopic scale, even though the resulting \textit{Gedankenexperiment} is
so far beyond the bounds of practicality as to challenge credulity (cf. Schr%
\"{o}dinger's own reference to his infamous cat scenario as "ridiculous")%
\footnote{%
Schr\"{o}dinger called \textit{Gedankenexperimenten} like that of his
eponymous cat "burleske F\"{a}lle". By contrast the Bohm-EPR \textit{%
Gedankenexperiment} famously leant itself to implementation as an \textit{%
actual} experiment with profound results for quantum nonlocality.}. So
consider a pair of spatially separated macroscopic systems $a,b$, composed
of $N\sim 10^{23}$ spin $%
{\frac12}%
$ particles each, in an entangled spin state $\left\vert \psi
_{Mac}\right\rangle $ of total $z$-component of angular momentum zero%
\begin{equation}
\left\vert \psi _{Mac}\right\rangle =\frac{1}{\sqrt{2}}\left[ \left\vert
\left( N\uparrow \right) \left( N\downarrow \right) \right\rangle
-\left\vert \left( N\downarrow \right) \left( N\uparrow \right)
\right\rangle \right]
\end{equation}%
Applied to this scenario, the EPR\ reasoning would lead one to conclude that
in $\left\vert \psi _{Mac}\right\rangle $ each of $a,b$ has a definite
macroscopic $z$-spin that quantum mechanics fails to describe. To reach this
conclusion, the argument would consider a hypothetical measurement of the $z$%
-spin on $a$ (a macroscopic object) and its outcome (a macroscopic value for
the $z$-spin on $a$) to conclude---\textit{independent of any measurements}%
---that $b$ has a definite macroscopic value of $z$-spin in state $%
\left\vert \psi _{Mac}\right\rangle $. Note that this argument need \textit{%
not} involve consideration of measurements of any other (incompatible)
spin-component on $a$.

However elegant the argument, EPRs conclusion is now generally taken to be
mistaken, primarily because of Bell's work and the subsequent experimental
violation of his eponymous inequalities. Now if EPR's argument is valid but
not sound, which of their assumptions are false? Even after Bohr's prompt
refutation and extensive more recent discussions of quantum nonlocality, I
believe there is still no consensus on exactly how to answer that question.
But I\ think many would follow Gisin$^{(30)}$ in pinning the blame on EPR's
locality assumptions, taking the failure of quantum mechanics to satisfy all
these assumptions to show why some of its (verified) predictions violate
Bell inequalities derived from them. More specifically, the condition Shimony%
$^{(31)}$ called Outcome Independence fails for quantum mechanics, as
illustrated by the fact that the quantum mechanical probability for outcome $%
\eta $ of a measurement of $\varphi $-spin on $a$ in the Bohm EPR scenario
conditional on a measurement of $\varphi ^{\prime }$-spin on $b$ depends on
the outcome $\eta ^{\prime }$ of the latter measurement (though it does not
depend on \textit{which} $\varphi ^{\prime }$-spin-component (if any) is
measured on $b$ if the outcome of any such measurement is ignored). This
failure of Outcome Independence does not facilitate signalling between
spacelike separated locations, and a variety of proofs have been offered
that quantum mechanical nonlocality is innocuous because it does not permit
such superluminal signalling.

But, as Maudlin$^{(32)}$ pointed out, there is still a problem reconciling
quantum mechanical nonlocality with relativity. Recall that, according to
the EPR\ argument, a measurement on $a$ projects $\left\vert \psi
_{s}\right\rangle $ onto the state $\left\vert \psi ^{\prime }\right\rangle $
in which the $\varphi $-spin of $b$ is definite. EPR took this to be a
straightforward application of quantum mechanics itself, unlike their
reality criterion and locality assumptions which were motivated by more
general physical considerations. If quantum mechanical description is
complete, in conflict with EPR's conclusion, then the $\varphi $-spin of $b$
was \textit{not }definite prior to the measurement on $a$. But if the $a,b$
measurements are spacelike separated events, then they have no invariant
temporal order, and any attempt to specify the spacetime location at which
the $\varphi $-spin of $b$ becomes definite must appeal to structure not
provided by a relativistic space-time and hard to reconcile with it.

Despite this problem, Lalo\"{e}$^{(12)}$ takes the analysis of his BEC 
\textit{Gedankenexperiment} to predict that spin-component measurements on a
few microscopic particles in $A$ will \textit{immediately} create a
spontaneous transverse polarization of a macroscopic assembly of spins in $B$%
.

\begin{quote}
what standard quantum mechanics describes here is not something that
propagates along the state and has a physical mechanism... it is just
'something with no time duration' that is a mere consequence of the
postulate of quantum measurement (wave packet reduction). (p.45)
\end{quote}

In fact the analysis he has given does not even establish the claim that,
following these measurements in $A$, a \textit{single} \textit{measurement}
of the \textit{total} $\varphi $-spin in $B$ would (almost) certainly yield
the predicted, definite macroscopic outcome. For that analysis concerns only
multiple (successful) \textit{microscopic} measurements of $\varphi $-spin
on individual particles in specific tiny regions of $B$. But it is true that
successfully measuring the $\varphi $-spin of each of a macroscopic number
of particles within $B$ and adding the results would be \textit{one} way of
measuring (a significant portion of) the total $\varphi $-spin in $B$.
Moreover, one can show that the expectation value of total $\varphi $-spin
in $B$ will be macroscopic after even a \textit{single} microscopic
transverse spin measurement in $A$. So it would be very surprising if an
extension of Lalo\"{e}'s$^{(12)}$ analysis did \textit{not} establish this
claim.

How does the EPR argument apply to Lalo\"{e}'s$^{(12)}$ BEC \textit{%
Gedankenexperiment}, in which the $a,b$ condensates overlap only in remote
regions $A,B$, a few $m\sim 100$ successful transverse spin-component
measurements are performed in $A$, and a macroscopic number of particles is
present in $B$? Here is what he says

\begin{quote}
We have a situation that is similar to the usual EPR\ situation:
measurements performed in $A$ can determine the direction of spins in both
regions $A$ and $B$. If we rephrase the EPR argument to adapt it to this
case, we just have to replace the words 'before the measurement in $A$' by
'before the series of measurements in $A$', but all the rest of the
reasoning remains exactly the same: since the elements of reality in $B$
cannot appear under the effect of what is done at an arbitrary distance in
region $A$, these elements of reality must exist even before the
measurements performed in $A$. Since the double Fock state $\left( \ref%
{double Fock}\right) $ of quantum mechanics does not contain any information
on the direction of spins in $B$, this theory is incomplete. (p.46)
\end{quote}

There is one clear disanalogy between the Bohm-EPR scenario and Lalo\"{e}'s$%
^{(12)}$ BEC \textit{Gedankenexperiment}. Even if an individual
spin-component measurement is projective, the sequence of measurements
performed in $A$ does not collapse the state $\left( \ref{double Fock}%
\right) $ into an eigenstate of total $\varphi $-spin in $B$: at most it
produces a state of the BECs for which a measurement of total $\varphi $%
-spin in $B$ is \textit{very likely} to give a particular result. Hence the
EPR\ reality criterion cannot be applied as stated, since it specifies
probability \textit{unity}. This disanalogy does not appear for the
macroscopic Bohm-EPR\ state $\left\vert \psi _{Mac}\right\rangle $, which is
in this respect a better macroscopic generalization of the original
Bohm-EPR\ state $\left\vert \psi _{s}\right\rangle $. Does this disanalogy
matter? I\ think it does.

Since they are arguing that quantum mechanical description is incomplete,
EPR need to have in mind a clear rival view of what it would be for it to be
complete. Quantum mechanics represents the (pure) state of systems by a
wave-function or state vector: how could such a mathematical object be
considered to offer a complete description of a system's properties? A
natural answer is that given by the so-called eigenvalue-eigenstate link:
observable $O$ represented by self-adjoint operator $\hat{O}$ has value $%
o_{i}$ on a system if and only if the state of that system can be
represented by pure state $\left\vert \omega _{i}\right\rangle $ where $\hat{%
O}\left\vert \omega _{i}\right\rangle =o_{i}\left\vert \omega
_{i}\right\rangle $. Indeed, EPR apply this link in both directions in
section 1\ of their paper. To adapt EPR's reasoning to Lalo\"{e}'s$^{(12)}$
BEC \textit{Gedankenexperiment} one would have to modify it to avoid relying
on the eigenvalue-eigenstate link.

Einstein's own preferred variant on EPR\footnote{%
For which, see Einstein ($(33)$, pp. 340-2; $(34)$, pp. 320-24; $(35)$, pp.
82-87).}\ does \textit{not }rely on the eigenvalue-eigenstate link. Instead
it directly argues for incompleteness of description by the wave-function.
As applied to the Bohm-EPR scenario Einstein$^{(34)}$ would reason that
while a measurement of $z$-spin on $a$ would collapse $b$'s state onto an
eigenstate of $z$-spin, a measurement of $x$-spin on $a$ would collapse the
state of $b$ onto an eigenstate of $x$-spin. By locality ("Grundsatz II" of
Einstein$^{(34)}$), neither measurement could influence the real state of $b$%
, which would therefore be the same no matter what measurement (if any) were
performed on $a$. But there is no way to understand both an $x$-spin
eigenstate and a $z$-spin eigenstate as offering a complete description of
the \textit{same} real state of $b$, since these eigenstates imply radically
different statistical predictions for the results of measurements on $b$.

Einstein's preferred mode of reasoning cannot be applied directly to Lalo%
\"{e}'s$^{(12)}$ BEC \textit{Gedankenexperiment}. For even if the $m$
transverse spin measurements on particles in $A$ are projective, they do not
project the quantum state onto a pure state that has the form of a tensor
product, one factor of which has support confined to $B$ and so could be
taken to describe just the contents of $B$.

Both the reasoning of the EPR\ argument and that of Einstein's preferred
variant \textit{may}, however, be readily applied to the macroscopic
generalization of Bohm-EPR represented by the state $\left\vert \psi
_{Mac}\right\rangle $, provided only that one takes the conclusion to be the
incompleteness of the description offered by $\left\vert \psi
_{Mac}\right\rangle $ of the real state of $b$ in that scenario. On the
other hand, Lalo\"{e}'s$^{(12)}$ BEC \textit{Gedankenexperiment }has the
distinct advantage of not being totally beyond the bounds of practicality.
As he says, progress in experimental studies of dilute gas BECs may bring us
within reach of producing systems of condensates for whose quantum
mechanical modeling the double Fock state provides a reasonable
idealization, and whose temporal evolution does not render an analysis in
terms of multiple simultaneous measurements wholly irrelevant. Just as the
Bohm-EPR scenario is no longer merely a \textit{Gedankenexperiment}, we may
be on the verge of realizing variants of Lalo\"{e}'s$^{(12)}$ BEC \textit{%
Gedankenexperiment} as real experiments.

\section{What Bohr would (or should?) have said}

Lalo\"{e}$^{(12)}$ presents his \textit{Gedankenexperiment} as a challenge
to Bohr's$^{(36)}$ refutation of the EPR\ argument in these words:

\begin{quote}
What is new here is that the EPR\ elements of reality in $B$ correspond to a
system that is macroscopic. One can no longer invoke its microscopic
character to deprive the system contained in $B$ of any physical reality!
The system can even be at our scale, correspond to a macroscopic
magnetization that can be directly observable with a hand compass; is it
then still possible to state that it has no intrinsic physical reality? When
the EPR\ argument is transposed to the macroscopic world, it is clear that
Bohr's refutation does not apply in the form written in his article; it has
to be at least modified in some way. (pp. 46-7)
\end{quote}

On the contrary, I venture that Bohr's reasoning in his refutation of EPR
applies equally well to Lalo\"{e}'s$^{(12)}$ \textit{Gedankenexperiment}. I
say "venture" rather than "claim" since any analysis based on an
interpretation of just what Bohr meant in his refutation must remain
tentative. What follows may, with some justification, be considered an
attempt to put words into Bohr's mouth that he would never have let pass his
lips!

In his refutation, Bohr$^{(36)}$ charged their reality criterion with fatal
ambiguity. The key passage is notoriously obscure, so I quote it at length.

\begin{quotation}
Of course there is in a case like that considered no question of a
mechanical disturbance of the system under investigation during the last
critical stage of the measuring procedure. But even at this stage there is
essentially the question of \textit{an influence on the very conditions
which define the possible types of predictions regarding the future behavior
of the system}. Since these conditions constitute an inherent element of the
description of any phenomenon to which the term 'physical reality' can
properly be attached, we see that the argumentation of the mentioned authors
does not justify their conclusion that quantum-mechanical description is
essentially incomplete. (p.700)
\end{quotation}

Note that Bohr here appears to \textit{deny} that the flaw in the argument
is the falsity of EPR's locality assumptions, while pointing to a different,
and perhaps deeper, problem with their assumptions about physical reality.
The problem is deeper in so far as the falsity of \textit{these} assumptions
would undermine the \textit{applicability} of notions of locality that rest
on them. Note also that in the quoted passage Bohr does not mention any
division between microscopic and macroscopic systems. Why, then would Lalo%
\"{e} suppose that his refutation is based on the denial of any physical
reality for microscopic systems that would not apply equally to \textit{%
macroscopic} systems?

\qquad The key phrase is surely that which Bohr himself stresses, namely "%
\textit{the very conditions which define the possible types of predictions
regarding the future behavior of the system"}. What does Bohr think those
conditions are? I\ believe a close reading of the rest of his reply to EPR
shows that what he has in mind here are the experimental conditions set up
by an experimenter who wishes to perform the measurement in question.
Moreover, this reading receives support from others of Bohr's writings. Bohr
would insist that any consideration even of a hypothetical measurement must
be based on some specification of the experimental conditions in order to be
well-grounded enough to play a role in an argument like that of EPR. His
idea is that since any ascription of physical reality to a quantity is
meaningful only in a well-defined experimental context, the element of
reality EPR argue for cannot be detached from the (hypothetical)
experimental context in which it is initially inferred to play an
independent role in the argument, including its conclusion.

Bohr maintained that the experimental conditions must be specified in
ordinary language, suitably enriched with the vocabulary of classical
physics. In his words (Bohr$^{(37)}$ )

\begin{quotation}
it is decisive to recognize that \textit{however far the phenomena transcend
the scope of classical physical explanation, the account of all evidence
must be expressed in classical terms.} The argument is simply that by the
word 'experiment' we refer to a situation where we can tell others what we
have done and what we have learned and that, therefore, the account of the
experimental arrangement and of the results of the observations must be
expressed in unambiguous language\ with suitable application of the
terminology of classical physics. (p.209)

The main point here is the distinction between the \textit{objects} under
investigation and the \textit{measuring instruments} which serve to define,
in classical terms, the conditions under which the phenomena appear. (pp.
221-2)
\end{quotation}

It has often been assumed that the distinction to which Bohr refers here is
one based on \textit{size}: that the apparatus is macroscopic, and so
accurately treatable by classical physics, while the object under
investigation is microscopic, and so must be treated quantum mechanically.
That this assumption is false becomes apparent when one reads the discussion
in Bohr$^{(37)}$ of his debates with Einstein, in which he applies quantum
mechanics to macroscopic objects without even feeling the need to comment on
the fact.

The distinction Bohr has in mind is a \textit{pragmatic }one: in order to
apply quantum mechanics to a system at all, he believes, one must describe
the whole experimental arrangement surrounding that system \textit{%
classically}. That is true whether the system under investigation is
microscopic or macroscopic. But any object that one (perforce) described
classically when it figured in the experimental arrangement for
investigating some other system may \textit{itself} be made the system under
investigation in the context of a \textit{different} experimental
arrangement, in which case it would be legitimate to apply quantum mechanics
to it in that context, and even necessary if classical physics proved
inadequate to predict results of observing it in that context.

Bohr's response to EPR did not rest on the denial of physical reality to
isolated microscopic systems. Instead, it rested on the denial of
context-independent attributions of properties (or rather physical
quantities) to systems of \textit{any} size. He took this denial to be
required by the transition to quantum physics from classical physics. As he
put it,

\begin{quote}
The necessity of discriminating in each experimental arrangement between
those parts of the physical system considered which are to be treated as
measuring instruments and those which constitute the objects under
investigation may indeed be said to form a \textit{principal distinction
between classical and quantum-mechanical descriptions of physical phenomena}%
. ($(36)$, p.701)
\end{quote}

On my reading, this passage makes clear just how radical Bohr's view of
quantum mechanical description was. In his view, with quantum mechanics, all
ascriptions of physical reality to properties of systems become contextual:
taken out of context, they lack significance. This denial of significance
rests on a pragmatist rather than a verificationist view of meaning\footnote{%
Here an analogy may be helpful. Consider the following message carved on a
tree-trunk deep in a forest: "I'll meet you here tomorrow". This message is
significant only in a context which specifies to whom 'I' and 'you' refer
and on what day the message was carved. Absent such a context, the message
is useless and so lacks pragmatic significance even though its semantic role
in English is perfectly clear. Supplying the context renders the message
verifiable.}. To supply the context needed to render meaningful the
ascription of a property to a system that one has decided to treat as "an
object under investigation" when applying quantum mechanics, one must
describe \textit{other} surrounding systems classically. This does not mean
that those other systems \textit{are} classical rather than quantum
mechanical. Still less does it mean that there is a special class of systems
("the macroscopic systems") which must be described classically. But it does
mean that there can be no purely quantum mechanical description of the
world, or even of any part of the world to which one contemplates applying
quantum mechanics. Lalo\"{e}'s$^{(12)}$ BEC \textit{Gedankenexperiment}
helps to bring out this radical character of Bohr's view even though it does
not challenge it. On Bohr's view, once one has decided to apply quantum
mechanics to the system of BECs in this \textit{Gedankenexperiment,} even
the ascription of a macroscopic magnetization to part of that system in a
region lacks significance, absent classically described conditions external
to the system. It may be hard to accept, but it is no refutation of this
view, that bringing up a hand compass renders that ascription not only
meaningful but true.

This response to Lalo\"{e}'s$^{(12)}$ BEC \textit{Gedankenexperiment} has
interesting implications for the claim that classical physics (or parts of
it, including classical mechanics) is reducible to quantum physics
(including quantum mechanics). If one endorses the response, then one has
strong grounds for denying such claims of reducibility. According to
classical physics, the behavior of a hand compass near region $B$ containing
enough of the BEC system following $100$ or so measurements of transverse
spin on particles in region $A$ would warrant ascribing a macroscopic
magnetization to (the contents of) that region. Any reduction of classical
physics to quantum physics here would involve showing that this is true%
\footnote{%
One might qualify this with 'approximately', but the qualitative nature of
the ascription renders this irrelevant.}. But if one endorses (what I take
to be) Bohr's response, this is \textit{not} true, or even significant,
outside of an appropriate context. So the most one could expect is a \textit{%
contextual} reduction of classical to quantum physics. But even this would
elude one in so far as at some stage the assumed \textit{context} could not
be described within a significant application of quantum physics\footnote{%
This train of thought may be what Landau and Lifshitz$^{(38)}$ had in mind
when they said
\par
\begin{quote}
"quantum mechanics occupies a very unusual place within physical theories:
it contains classical mechanics as a limiting case, yet at the same time it
requires this limiting case for its own formulation." p.3
\end{quote}
}.

\section{Emergent Properties and Emergent Objects}

Lalo\"{e} summarizes the essence of his $(12)$ as follows

\begin{quote}
in some quantum situations where macroscopic systems populate Fock states
with well defined populations, the EPR\ argument becomes significantly
stronger than in the historical example with two microscopic particles. The
argument speaks eloquently in favour of a pre-existing relative phase of the
two states ... but certainly not in favour of the orthodox point of view
where the phase appears during the measurements. \ (p.51)
\end{quote}

In spite of the objections I have lodged against his argument, Lalo\"{e}\
here expresses an important insight that should not be lost if we wish to
understand the emergence of relative phase between BECs. The EPR\ argument
was directed against a popular version of the Copenhagen interpretation that
takes the quantum state to describe intrinsic properties of a system it
represents, and measurement to project the quantum state onto a new one that
describes the system's new intrinsic properties. If one tries to understand
the emergence of relative phase between BECs initially in a double Fock
state as a stochastic dynamical process mediated by successive projective
measurements on individual particles in the condensate, then, as Lalo\"{e}
goes on to say, surprising non-local effects appear in the macroscopic world
(which, I\ might add, are extremely difficult to reconcile with relativistic
spacetime structure, even though they do not admit superluminal signalling).

Must one who rejects ths popular version of the Copenhagen interpretation
conclude that the relative phase between BECs was definite already prior to
measurements on its constituent particles, which simply progressively reveal
that pre-existing phase $\Lambda $? Drawing this conclusion on the basis of
EPR-type reasoning, one would take $\Lambda $ to be an \textit{additional}
variable characterizing the BECs in quantum state $\left( \ref{double Fock}%
\right) $, initially hidden but gradually revealed by transverse spin
measurements. But further investigations by Lalo\"{e} and Mullin$^{(10)}$
effectively block this route. They derive
(Bell-Clauser-Horne-Shimony-Holt)-type inequalities for carefully chosen
observables of particles in quantum state $\left( \ref{double Fock}\right) $
on the assumption of a pre-existing relative phase between the condensates,
and show that quantum mechanics predicts their violation in that state.%
\footnote{%
It is, however, noteworthy that experimental violation would be
extraordinarily difficult to arrange because it would be essential to
perform measurements on \textit{all the particles} in the system of
condensates.} So, just as in the Bohm-EPR case, the intended conclusion of
EPR-type reasoning here proves to be incompatible with quantum mechanics
itself.

There is a different way to use a pre-existing relative phase $\Lambda $ to
account for the interference exhibited by a system of two similar
condensates as a result of transverse spin measurements. It is to \textit{%
deny} that their initial quantum state is correctly represented by $\left( %
\ref{double Fock}\right) $, and to claim that it is rather a phase state $%
\left( \ref{phase state}\right) $. The analysis of section 4 shows that
these two quantum states lead to identical interference patterns for the
phenomena considered there. Of course, by taking this line one is evading
rather than solving the problem of understanding how a relative phase
emerges in the double Fock state $\left( \ref{double Fock}\right) $. Such
evasion could be justified by an argument as to why any natural preparation
procedure for a system of condensates of the type we have been considering
would give rise to the phase state $\left( \ref{phase state}\right) $
instead. But if one recalls that the whole discussion of interference
between similar BECs was provoked by experiments like those of Andrews 
\textit{et. al.}$^{(2)}$, the prospects of developing such an argument seem
bleak.

Leggett$^{(21)}$, for example, says this

\begin{quote}
The authors start with a trap which is split into two by a laser-induced
barrier so high that the single-atom tunnelling time between the two wells
is greater than the age of the universe. They then condense clouds of $^{87}$%
Rb atoms independently in the two wells and allow them to come to thermal
equilibrium. At this point there seems no doubt that the correct quantum
mechanical wave-function of the system is, schematically, [of the form of a
double Fock state] (p.138)
\end{quote}

He goes on to show that the time-evolution of each component to bring them
into overlap after removal of the laser barrier will not make this double
Fock state approach a phase state. Lalo\"{e}$^{(12)}$ argues that
environmental decoherence can favor phase states over double Fock states,
but dismisses this as a reason to reject his analysis in terms of double
Fock states. Since coupling with the environment tends to produce an
improper \textit{mixture} of phase states, if there is any interference in a
system like that analyzed, this cannot be accounted for by appeal to a 
\textit{pure} phase state of the BEC system.

It is interesting to contrast the case of a system of dilute gas BECs in a
double Fock state $\left( \ref{double Fock}\right) $ with other systems
involving a pair of condensates that exhibit interference phenomena
accounted for by appeal to a relative phase between them. When a pair of
conductors separated by a thin metal oxide junction is cooled to become
superconducting, a current flows across the junction even in the absence of
an applied voltage. This DC\ Josephson effect may be explained quantum
mechanically by appeal to a well-defined phase difference $\phi $ across the
junction in the wave-function representing the state of the system: the DC\
current is proportional to $\sin \phi $. Leggett and Sols$^{(25)}$ write the
wave-function as follows,

\begin{equation}
\Phi \sim \left( \left\vert a\right\vert e^{i\phi /2}\psi _{L}+\left\vert
b\right\vert e^{-i\phi /2}\psi _{R}\right) ^{N}  \label{Josephson}
\end{equation}%
where the system consists of $N$ "bosons" (Cooper pairs) and $\psi _{L}$ ($%
\psi _{R}$) is the Schr\"{o}dinger amplitude for a boson to be on the left
(right) of the junction. Note the analogy with the phase state $\left( \ref%
{phase state}\right) $\footnote{%
Leggett and Sols (1991) actually apply their analysis to a generic Josephson
effect in a Bose superfluid, of which a superconductor is one example,
another being superfluid Helium.}. If positing a phase state like $\left( %
\ref{Josephson}\right) $ is indeed the only way to explain the Josephson
effect, then just as in the case of the dilute gas BECs, one should ask how
a relative phase emerges. One possible answer is that there is \textit{always%
} some relative phase difference between any pair of similar superconductors
(even those prepared independently and arbitrarily far away from each
other), and its (random) value emerges as a result of spontaneous breaking
of the $U(1)$ symmetry. Leggett and Sols$^{(25)}$ reject this answer, and
Leggett$^{(26)}$ advances an interesting argument for doing so.

To set the context for this argument, note that the state $\left( \ref%
{Josephson}\right) $ may be expanded in a basis of double Fock states $%
\left\vert N_{a},N_{b}\right\rangle $ as%
\begin{equation}
\Phi \sim \sum\limits_{M=-N/2}^{+N/2}\left\vert C_{M}\right\vert e^{iM\phi
}\left\vert N_{a},N_{b}\right\rangle
\end{equation}%
where $\left( N_{a}+N_{b}\right) =N$ and%
\begin{equation}
\left\vert N_{a},N_{b}\right\rangle \sim \hat{a}\dag _{\psi _{L}}^{N_{a}}%
\hat{a}\dag _{\psi _{R}}^{N_{b}}\left\vert 0\right\rangle
\end{equation}%
It follows that in state $\left( \ref{Josephson}\right) $ the difference
between the number of bosons in the left-hand condensate and the number in
the right-hand condensate is indeterminate, even though together they
contain exactly $N$ bosons. This may be acceptable in the typical situation
in which one takes $\left( \ref{Josephson}\right) $ to represent the state
of a pair of similar condensates, prepared together and spatially separated
only by a thin junction. But it is harder to stomach if the left and right
hand condensates have been separately prepared in different continents!

Leggett$^{(26)}$ rejects this \textit{outr\'{e}} suggestion, and presents a
thought experiment as a reason for doing so.

\begin{quote}
The "experiment" simply consists in weighing them at separate times ... that
can be arbitrarily far separated, so as to determine the number difference $%
\left[ N_{a}-N_{b}\right] $ at these times, without ever making Josephson
contact between them. (p.459)
\end{quote}

If $\left( \ref{Josephson}\right) $ correctly represents their total state,
then there is no reason to expect the results to agree: indeed, one would
expect them to differ by an amount of the order of $N^{%
{\frac12}%
}$. If, on the other hand, the correct representation is a double Fock state
(or mixture of these), then the results would be expected to agree (within
the margin of error of the experiment). Leggett$^{(26)}$ concludes

\begin{quote}
I can see no reason whatever to doubt that it is this latter conclusion
which would be found experimentally, so that in \textit{this} (operationally
defined) sense, the statement that "two superfluids which have never seen
one another before nevertheless have a definite relative phase" is, I
believe, false. (\textit{ibid.})
\end{quote}

But suppose we take spontaneous symmetry breaking absolutely seriously here
and consider what we should say if the results of Leggett's thought
experiment were to confound his firm expectations. In that case, I submit,
we should have evidence for more than just the emergence of relative phase
in BECs consequent upon spontaneous symmetry breaking: we should have reason
to accept the spontaneous emergence of composite \textit{objects}---the BECs
themselves.

Here we have at least a conceptual possibility not (to my knowledge)
contemplated by philosophers interested in emergence. When philosophers have
considered the possibility of emergent objects, they have had in mind a case
in which an object composed of a perfectly determinate set of microscopic
parts possesses an emergent \textit{property} (however that notion is
analyzed).\footnote{%
See, for example, section 1 of Bedau$^{(15)}$.} But what we are presently
contemplating is a case in which each of two objects, composed of nothing
but microscopic parts of a certain kind, contains no definite \textit{number}
of these objects, although together the \textit{pair} is composed of a
definite number of these constituent parts.

At first this may seem analogous to more familiar cases:\ consider a cat's
tail and the rest of its body, Siamese twins, two colliding galaxies, or the
stratosphere and troposphere. But in such cases the parts of the total
system are spatially contiguous and the indeterminateness of composition of
each where they join is naturally attributed to the vagueness of the
language we use to describe them. If two similar BECs, independently
prepared on different continents, had a definite relative phase, then each
BEC would be an emergent object in a much stronger sense. The
indeterminateness of composition could not be localized to any spatially
intermediate region and would be distributed equally among all their
component bosons. Perhaps we have here a new candidate for the
metaphysician's disputed category of vague objects?\pagebreak

{\LARGE References}

1. Anderson, M.H. \textit{et. al.}: Observation of Bose-Einstein
condensation in a dilute atomic vapor. \textit{Science 269}, 198-201 (1995)

2. Andrews, M.R., \textit{et. al.}: Observation of interference between two
Bose condensates. \textit{Science 275}, 637-641 (1997)

3. Laughlin, R.B. and Pines, D.: The theory of everything. \textit{%
Proceedings of the National Academy of Sciences 97}, 28-31\textit{\ }(2000)

4. Morrison, M.: Emergence, reduction, and theoretical principles:
Rethinking fundamentalism. \textit{Philosophy of Science 73}, 876-87 (2006)

5. Javanainen, J. and Yoo, S.M.: Quantum phase of a Bose-Einstein condensate
with an arbitrary number of atoms. \textit{Physical Review Letters 76},
161-4 (1996)

6. Castin, Y. and Dalibard, J.: Relative phase of two Bose-Einstein
condensates. \textit{Physical Review A55}, 4330-7 (1997)

7. Lalo\"{e}, F.: The hidden phase of Fock states; quantum non-local
effects. \textit{The European Physics Journal D 33}, 87-97 (2005)

8. Mullin, W.J., Krotkov, R. and Lalo\"{e}, F.: Evolution of additional
(hidden) quantum variables in the interference of Bose-Einstein condensates. 
\textit{Physical Review A74}, 023610:1-11 (2006)

9. Mullin, W.J., Krotkov, R. and Lalo\"{e}, F.: The origin of the phase in
the interference of Bose-Einstein condensates. \textit{American Journal of
Physics 74}, 880-87 (2006)

10. Lalo\"{e}, F. and Mullin, W.J.: Einstein-Podolsky-Rosen argument and
Bell inequalities for Bose-Einstein spin condensates. \textit{Physical
Review A77}, 022108:1-17 (2008)

11. Paraoanu, G.S.: Localization of the relative phase \textit{via}
measurements. \textit{Journal of Low Temperature Physics 153}, 285-93 (2008)

12. Lalo\"{e}, F.: Bose-Einstein condensates and EPR quantum non-locality.
In: T.M. Nieuwenhuizen \textit{et. al. }(eds.) \textit{Beyond the Quantum},
pp. 35-52. World Scientific, Singapore (2007)

13. Weinberg, S.: \textit{Dreams of a Final Theory.} Random House, New York
(1992)

14. McLaughlin, B.:\ Emergence and supervenience. \textit{Intellectica 25},
25-43 (1997)

15. Bedau, M.: Downward causation and autonomy in weak emergence. \textit{%
Principia Revista Internacional de Epistemologica 6}, 5-50 (2003)

16. Humphreys, P.: How properties emerge. \textit{Philosophy of Science 64},
1-17 (1997)

17. Teller, P.: A contemporary look at emergence. In: Beckerman, A., Flohr,
H., and Kim, J. (eds.) \textit{Emergence or Reduction? Essays on the
Prospects of Nonreductive Physicalism}, pp. 139-53. de Gruyter, Berlin (1992)

18. Anderson, P.W.: More is different. \textit{Science 177}, 393-396 (1972)

19. Weinberg, S.: \textit{The Quantum Theory of Fields II}. Cambridge
University Press, Cambridge (1996)

20. Wilson, M.: \textit{Wandering Significance}.\textit{\ }Oxford University
Press, Oxford (2006)

21. Leggett, A.J.: \textit{Quantum Liquids}. Oxford University Press, Oxford
(2006)

22. Ruetsche, L.: Johnny's so long at the ferromagnet.\ \textit{Philosophy
of Science 73 [Proceedings]}, 473-486 (2006)

23. Goldstone, J., Salam, A. and Weinberg, S.: Broken Symmetries. \textit{%
Physical Review 127}, 965-70 (1962)

24. Streater, R.: The Heisenberg ferromagnet as a quantum field theory. 
\textit{Communications in Mathematical Physics 6}, 233-47 (1967)

25. Leggett, A.J. and Sols, F.: On the concept of spontaneously broken gauge
symmetry in condensed matter physics. \textit{Foundations of Physics 21},
353-64 (1991)

26. Leggett, A.J.: Broken gauge symmetry in a Bose condensate. In: Griffin,
A., Snoke, D.W., and Stringari, S. (eds.) \textit{Bose-Einstein Condensation}%
, pp. 452-62. Cambridge University Press, Cambridge (1995)

27. Teller, P.: Relational holism and quantum mechanics.\ \textit{British
Journal for the Philosophy of Science 37}, 71-81 (1986)

28. Leggett, A.J.: Topics in the theory of the ultracold dilute alkali
gases. \textit{Modern Physics Letters B14 (Supplementary Issue)}, 1-42 (2000)

29. Einstein, A., Podolsky, B. and Rosen, N.: Can quantum-mechanical
description of physical reality be considered complete? \textit{Physical
Review 47}, 777-80 (1935)

30. Gisin, N.: Non-realism: deep thought or a soft option? arXiv:0901.4255v2
[quant-ph] 18 Aug 2009

31. Shimony, A.: Events and processes in the quantum world. In: Penrose, R.,
and Isham, C. (eds.) \textit{Quantum Concepts in Space and Time}, pp.
182-203. Oxford University Press, Oxford (1986)

32. Maudlin, T.: \textit{Quantum Non-locality and Relativity}. Blackwell,
Oxford (1994)

33. Einstein, A.: Physik und Realit\"{a}t. \textit{Journal of the Franklin
Institute 221}, 313-347 (1936)

34. Einstein, A.: Quantenmechanik und Wirklichkeit, \textit{Dialectica 2},\
320-4 (1948)

35. Einstein, A.: Autobiographical Notes. In: Schilpp, P.A. (ed.) \textit{%
Albert Einstein:\ Philosopher-Scientist}, pp. 5-94. Open Court, La Salle,
Illinois (1949)

36. Bohr, N.: Can quantum-mechanical description of physical reality be
considered complete? \textit{Physical Review 48}, 696-702 (1935)

37. Bohr, N.: Discussion with Einstein on epistemological\ problems in
atomic physics. In: Schilpp, P.A. (ed.) \textit{Albert Einstein:\
Philosopher-Scientist}, pp. 201-41. Open Court, La Salle, Illinois (1949)

38. Landau, L. D. and Lifshitz, E. M.: \textit{Quantum Mechanics:
Non-Relativistic Theory,} \textit{3rd edition}. Pergamon, Oxford (1977)

\end{document}